\documentclass[letterpaper, 10 pt, conference]{IEEEtran}
\newcounter{mytempeqncnt}

\newcommand{\qed}{\nobreak \ifvmode \relax \else
      \ifdim\lastskip<1.5em \hskip-\lastskip
      \hskip1.5em plus0em minus0.5em \fi \nobreak
      \vrule height0.75em width0.5em depth0.25em\fi}

\ifCLASSINFOpdf
  \usepackage[pdftex]{graphicx}
  \graphicspath{{./}{Figs/}}
  \DeclareGraphicsExtensions{.pdf,.jpeg,.png}
\else
\fi
\usepackage{amssymb}
\usepackage{amsmath}
\usepackage{mathtools}
\usepackage{bigdelim}

\newcommand{\MeijerG}[7]{G^{#1,#2}_{#3,#4} \left(\! \begin{smallmatrix} #5 \\ #6 \end{smallmatrix} \middle\vert #7 \!\right) }

\IEEEoverridecommandlockouts

\begin{document}

\title{On the Capacity of the Underwater Acoustic Channel with Dominant Noise Sources}
\author{\IEEEauthorblockN{Mustafa A. Kishk and Ahmed M. Alaa}
\IEEEauthorblockA{Department of Electronics and Communications Engineering,\\
Faculty of Engineering, Cairo University,
Giza 12613, Egypt.\\ Emails: mustafa.kishk@gmail.com and aalaa@eece.cu.edu.eg}}
\maketitle
\begin{abstract}
This paper provides an upper-bound for the capacity of the underwater acoustic (UWA) channel with dominant noise sources and generalized fading environments. Previous works have shown that UWA channel noise statistics are not necessary Gaussian, especially in a shallow water environment which is dominated by impulsive noise sources. In this case, noise is best represented by the Generalized Gaussian (GG) noise model with a shaping parameter $\beta$. On the other hand, fading in the UWA channel is generally represented using an $\alpha$-$\mu$ distribution, which is a generalization of a wide range of well known fading distributions. We show that the Additive White Generalized Gaussian Noise (AWGGN) channel capacity is upper bounded by the AWGN capacity in addition to a constant gap of $\frac{1}{2} \log \left(\frac{\beta^{2} \pi e^{1-\frac{2}{\beta}} \Gamma(\frac{3}{\beta})}{2(\Gamma(\frac{1}{\beta}))^{3}} \right)$ bits. The same gap also exists when characterizing the ergodic capacity of AWGGN channels with $\alpha$-$\mu$ fading compared to the faded AWGN channel capacity. We justify our results by revisiting the sphere-packing problem, which represents a geometric interpertation of the channel capacity. Moreover, UWA channel secrecy rates are characterized and the dependency of UWA channel secrecy on the shaping parameters of the legitimate and eavesdropper channels is highlighted. 
            	
\end{abstract}
\IEEEpeerreviewmaketitle
\section{Introduction}
Vast efforts have been lately devoted to the analysis of the Underwater Acoustic (UWA) communications channel. Various applications that encounter an UWA channel include underwater sensor networks that are installed at the bottom of oceans for different purposes like data collection, pollution monitoring, assisted navigation, and offshore exploration [1]. Unmanned and autonomous underwater vehicles that carry sensors for natural underwater exploration are also applications that push towards studying the behavior of the UWA channel. Previous works have analyzed the different characteristics of the UWA channels in terms of error rate performance and receiver design [2]-[6]. In order to obtain the fundamental performance limits of the UWA channel, it is mandatory to characterize the statistical behavior of the additive noise and fading processes exhibited by transceivers operating in underwater environments. In [2], the authors studied different types of noise sources that exist in the underwater nature. Existence of all of these sources will lead to the standard Gaussian distributed noise [7]. However, in other cases, such as communications in shallow water levels, one of these sources dominate, which leads to a different noise model. The standard Gaussian noise model does not hold in dominant noise scenarios, and it was found that Generalized Gaussian (GG) distributions accurately describe the behavior of such channels. In [4] and [5], the error rate for the channels with GG noise was evaluated. The authors also studied the impact of designing optimal receivers based on AWGN channel while actually operating in AWGGN underwater environments. To the best of our knowledge, the capacity of the AWGGN channel was never evaluated before. Moreover, it is important to characterize the fading statistics in the underwater environment. In [8], many experiments have been done trying to define the best representation of the UWA channel fading. The experiments were carried out by transmitting continuous wave signals in shallow water at depth of 6 to 12 meters at {\it Rio de Janeiro}, {\it Brazil}. The received signals was analyzed at different frequencies after removing the noise effect, and it was found that the $\alpha$-$\mu$ distribution accurately describes the UWA channel fading statistics. 

In this paper, we derive an upper bound on the achievable capacity of the UWA acoustic channel based on a GG noise model. It is shown that the UWA channel capacity is bounded by the conventional AWGN channel capacity added to a constant gap that depends on the GG distribution shaping parameter $\beta$. Next, we develop an upper bound on the ergodic capacity of the UWA channel based on a generalized $\alpha$-$\mu$ distribution. The proposed upper bound is generic and depends on the parameters tuple ($\beta$,$\alpha$,$\mu$), thus it described a wide range of UWA communications settings in terms of depth, frequency, and dominance of noise sources. We interpret the analytical results by revisiting the sphere-packing problem, which serves as a geometric interpretation of channel capacity. Moreover, we characterize the UWA channel secrecy when the dominant noise sources at the legitimate and eavesdropper receivers are different. 

The rest of the paper is organized as follows. In section II, the GG and the $\alpha$-$\mu$ distributions will be represented. In section III, the AWGGN UWA channel capacity upper bound will be derived. Section IV presents an expression for the capacity upper bound with both GG noise and $\alpha$-$\mu$ fading. Knowing that secrecy rate in a channel with some eavesdropper depends mainly on the channel capacity for both source-destination channel and source-eavesdropper channel, we study secrecy rate when these channels have GG noise model in section V. Analytical results will be shown in section VI. The paper is concluded in Section VII.

\section{UWA Channel Model}

\subsection{Additive White Generalized Gaussian Noise (AWGGN)}

For the AWGGN channel, the received signal $Y_{i}$ at the $i^{th}$ time instant results from adding the transmitted signal $X_{i}$ to a GG noise signal $N_{i}$ as follows
\[Y_{i} = X_{i} + N_{i}.\]
 It is assumed that consecutive noise samples are uncorrelated. The probability density function (pdf) of a noise sample $N$ is given by the GG distribution [9] 
\begin{equation}
\label{1}
f_{N}(N) = \frac{\beta}{2 \gamma \Gamma(\frac{1}{\beta})} \mbox{exp}\left(\frac{-|N-\mu_{N}|^{\beta}}{\gamma}\right), 
\end{equation} 
where $\mu_{N}$ is the mean of the pdf which is typically equal to zero for a noise process, $\Gamma(.)$ is the gamma function, $\gamma$ and $\beta$ are both scale and shape parameters of the GG pdf. The variance of $N$ is given by $\sigma_{N}^{2} = \frac{\gamma^{2} \Gamma(\frac{3}{\beta})}{\Gamma(\frac{1}{\beta})}$. Based on the value of $\beta$, the GG distribution may converge to other known densities. For instance, the channel reduces to a standard AWGN channel for $\beta$ = 2. Figure 1 depicts the pdf of a GG distributed random variable. It is shown that as the value of $\beta$ decreases, the pdf becomes more confined around its mean value, which means that it becomes ``{\it more deterministic}". For $\beta$ = 0.1, the pdf of the GG random variable is close to an impulse located at the mean value, which implies that small values of $\beta$ correspond to less uncertainity about the value of the noise sample.  

\begin{figure}[!t]
\centering
\includegraphics[width=3.25in]{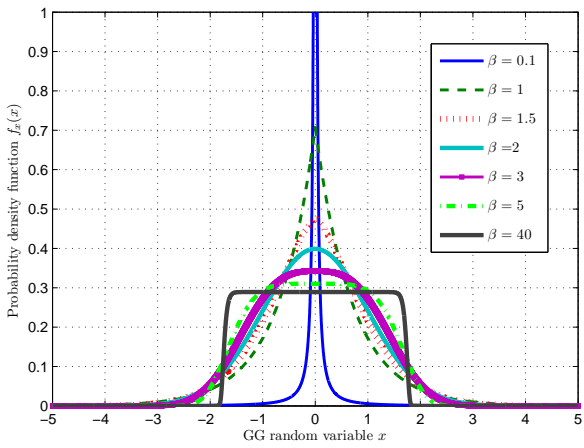}
\caption{The pdf of a GG distributed random variable for different values of $\beta$.}
\label{fig_sim}
\end{figure}

\subsection{The $\alpha$-$\mu$ fading distribution}

For the faded AWGGN channel, the received signal $Y_{i}$ at the $i^{th}$ time instant results from adding the transmitted signal $X_{i}$ mutliplied by a fading channel gain $h_{i}$  to a GG noise signal $N_{i}$ as follows
\[Y_{i} = h_{i} \, X_{i} + N_{i},\]
where coherent detection is assumed at the receiver. The pdf of $h$ is given by
\begin{equation}
\label{2}
f_{h}(h) = \frac{\alpha \mu^{\mu} h^{\alpha \mu-1}}{\tilde{h}^{\alpha \mu}\Gamma(\mu)} \mbox{exp}\left(-\mu \frac{h^{\alpha}}{\tilde{h}^{\alpha}} \right), 
\end{equation}
where $\alpha > 0$ is an arbitrary parameter, $\tilde{h} = \sqrt[\alpha]{\mathbb{E}\{h^{\alpha}\}}$, and $\mu = \frac{\mathbb{E}^{2}\{h^{\alpha}\}}{\mathbb{V}\{h^{\alpha}\}}$, given that $\mathbb{E}\{.\}$ and $\mathbb{V}\{.\}$ are the expectation and variance operators.     

\begin{figure}[!t]
\centering
\includegraphics[width=3.25in]{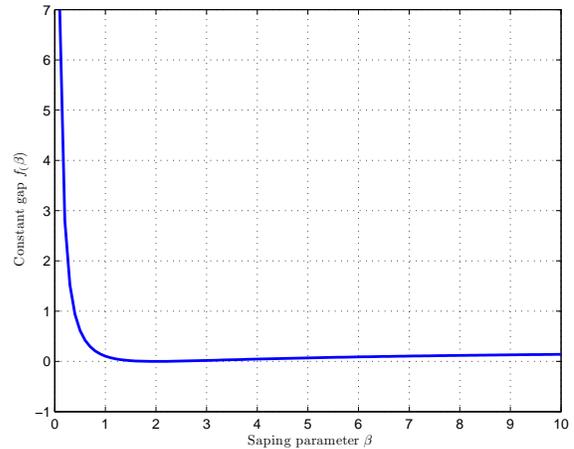}
\caption{The constant additive gap $f(\beta)$ versus the GG shaping parameter $\beta$.}
\label{fig_sim}
\end{figure}

\section{Capacity of the AWGGN Channel}
\subsection{Capacity Evaluation}
In order to calculate the capacity of the AWGGN channel, we start by deriving the differential entropy $h(.)$ of the GG random variable $N$. The differential entropy is given by
\begin{align}
\label{eqq}
h(N) &=& \mathbb{E}\{- \log(f_{N}(N))\} 
&=& - \int_{\mathbb{N}} f_{N}(N) \log(f_{N}(N)) dN,  
\end{align}
which can be easily evaluated as
\begin{equation}
\label{eqq2}
h(N) = \frac{1}{\beta} - \log\left(\frac{\beta}{2 \alpha \Gamma(\frac{1}{\beta})}\right) = \frac{1}{2} \log\left(\frac{4 e^{\frac{2}{\beta}}\alpha^{2} \left(\Gamma(\frac{1}{\beta})\right)^{2}}{\beta^{2}}\right).
\end{equation}

For some message signal $X$ and GG distributed noise $N$, the received signal $Y$ can be represented as $Y=X+N$. Assuming zero mean noise and message signal power $P$ then received power can be given by
\[\mathbb{E}\{Y^{2}\} = \mathbb{E}\{(X+N)^{2}\} = \mathbb{E}\{X^{2}\} + \mathbb{E}\{N^{2}\}.\]
Thus, the received signal power can be obtained by plugging the variance of the GG distribution as follows
\[\mathbb{E}\{Y^{2}\} = P + \frac{\gamma^{2} \Gamma(\frac{3}{\beta})}{\Gamma(\frac{1}{\beta})}.\]
By definition the channel capacity is the maximum mutual information $I(X;Y)$ which is given by:
\begin{align}
\label{eqq3}
I(X;Y) &= h(Y) - h(Y|X) \\
&= h(Y) - h(X + N|X) \\
&= h(Y) - h(N), 
\end{align}
where (7) follows from the fact that $h(X|X) = 0$. Following theorem 9.6.5 in [10], the differential entropy $h(Y)$ is bounded by the entropy of a gaussian signal as follows
\begin{equation}
h(Y) \leq \frac{1}{2} \log\left(2 \pi e \left(P+ \frac{\gamma^{2} \Gamma(\frac{3}{\beta})}{\Gamma(\frac{1}{\beta})}\right)\right).
\end{equation} 
By combining (4), (7), and (8), the capacity of the AWGGN channel $C_{AWGGN}$ is bounded by
\begin{equation}
C_{AWGGN} \leq \frac{1}{2} \log\left( \frac{\beta^{2} \pi e^{1-\frac{2}{\beta}}}{2 \alpha^{2} \left(\Gamma(\frac{1}{\beta})\right)^{2}} \left(P+ \frac{\gamma^{2} \Gamma(\frac{3}{\beta})}{\Gamma(\frac{1}{\beta})}\right) \right), 
\end{equation}   
where the bound is achieved only if the pdf of the input signal is chosen such that the received signal is gaussian, i.e. $f_{Y}(Y) = f_{N}(N) \star f_{X}(X) \sim \mathcal{N}(0, P)$, where $\star$ is the convolution operator. Taking into consideration that $C_{AWGGN}$ is lower bounded by $C_{AWGN}$, as the gaussian noise is a worst case (maximum entropy) noise [11]-[12], the capacity of the UWA channel when dominant noise sources exist forming a GG noise process with a shaping parameter $\beta$ is lower bounded by the AWGN capacity and upper bounded by the AWGN with an additive constant gap of $f(\beta) = \frac{1}{2} \log \left(\frac{\beta^{2} \pi e^{1-\frac{2}{\beta}} \Gamma(\frac{3}{\beta})}{2(\Gamma(\frac{1}{\beta}))^{3}} \right)$, for a noise variance of $\sigma_{N}^{2} = \frac{\alpha^{2} \Gamma(\frac{3}{\beta})}{\Gamma(\frac{1}{\beta}}$ as shown in eq. (10). Figure 2 depicts the constant capacity gap $f(\beta)$ versus the shaping parameter $\beta$. It is obvious that this gap is always positive, and is euqal to 0 only when $\beta$ = 2, which is the AWGN scenario. Thus, impulsive noise offers opportunities for improving capacity compared to AWGN, and the amount of capacity improvment is dependent on the value of $\beta$. Lower values of the shaping parameter corresponds to less entropy for the noise process, and thus more capacity. This conclusions fit with the results obtained in [11] and [12], where the authors proved that gaussian noise is a worst case additive noise, and the capacity of non-gaussian channels is bounded by the gaussian channel capacity.  
  
\begin{figure*}[!t]
\normalsize
\setcounter{mytempeqncnt}{\value{equation}}
\setcounter{equation}{9}
\begin{align}
\label{9}
\frac{1}{2} \log\left(1+\frac{P}{\sigma_{N}^{2}}\right) \leq C_{AWGGN} \leq \frac{1}{2} \log\left(1+\frac{P}{\sigma_{N}^{2}}\right) + \frac{1}{2} \log \left(\frac{\beta^{2} \pi e^{1-\frac{2}{\beta}} \Gamma(\frac{3}{\beta})}{2(\Gamma(\frac{1}{\beta}))^{3}} \right). 
\end{align}
\setcounter{equation}{\value{mytempeqncnt}+2}
\hrulefill
\vspace*{4pt}
\end{figure*} 

\subsection{Sphere-Packing Interpretation}
The analytical results obtained in the previous subsection can be interpreted geometrically by revisiting the sphere-packing problem. For an $K$-dimensional codeword $X$, the received codeword $Y = X + N$ where $N$ is the noise added, the received codeword in general lies with high probability in a sphere with volume $2^{K h(Y)}$ called the Y-sphere. For a certain codeword $X$, the received codeword $Y$ will lie in the noise sphere around this codeword $X$ with volume $2^{K h(Y|X)}$. Therefore, the maximum number of noise spheres $V$ that can be packed inside the Y-sphere without overlapping will be the ratio between their volumes $V = \frac{2^{K h(Y)}}{2^{K h(Y|X)}}$. In both cases of AWGN and AWGGN noise channels, $h(Y)$ have the same expression while the only difference is in the expression of the conditional differential entropy $h(Y|X)$, which is equal to $h(N)$. By revisiting the analysis in the previous subsection, it can be shown that
\[h(N_{AWGGN}) = h(N_{AWGN}) - f(\beta).\]
Thus, for the same noise variance, the number of GG noise spheres that can be packed into the Y-sphere is given by $V_{AWGGN} = \frac{2^{K h(Y)}}{2^{K h(N_{AWGN})-f(\beta)}} = 2^{f(\beta)} \, V_{AWGN}$. Therefore, for the same noise variance, AWGGN noise spheres have a volume that is less than the volume of the AWGN noise sphere with a factor of $\frac{1}{2^{f(\beta)}}$, which allows packing $2^{f(\beta)}$ more spheres in the GG channel. To sum up, when the noise spheres are {\it immersed} under water, they get {\it compressed} with a factor of $2^{f(\beta)}$, allowing to pack more noise spheres in the Y-sphere.  

\section{Capacity of the UWA channel with $\alpha$-$\mu$ fading}
A comprehensive experimental study in [8] has shown that the $\alpha$-$\mu$ fading distribution fits a wide range of underwater fading scenarios. Having obtained bounds on the AWGGN UWA channel capacity, we can conclude that the conditional capacity $C(\beta,\alpha,\mu | h)$ of the channel given a fading channel gain $h$ as
\[C(\beta,\alpha,\mu | h) \geq \frac{1}{2} \log\left(1+\frac{P \, h^{2}}{\sigma_{N}^{2}}\right),\]
and
\[C(\beta,\alpha,\mu | h) \leq \frac{1}{2} \log\left(1+\frac{P \, h^{2}}{\sigma_{N}^{2}}\right) + \frac{1}{2} \log \left(\frac{\beta^{2} \pi e^{1-\frac{2}{\beta}} \Gamma(\frac{3}{\beta})}{2(\Gamma(\frac{1}{\beta}))^{3}} \right). \]   
The ergodic capacity can be obtained by averaging the conditional capacity over the pdf of $h$ as follows
\[
C(\beta,\alpha,\mu) = \mathbb{E}_{h}\{C(\beta,\alpha,\mu | h)\}.
\] 
The bounds on the ergodic capacity can be easily obtained by solving the integral $\int_{h = -\infty}^{\infty} \log\left(1+\frac{P \,h^{2}}{\sigma_{N}^{2}}\right) \times \frac{\alpha \mu^{\mu} h^{\alpha \mu-1}}{\tilde{h}^{\alpha \mu}\Gamma(\mu)} \mbox{exp}\left(-\mu \frac{h^{\alpha}}{\tilde{h}^{\alpha}} \right) dh$, which corresponds to the ergodic capacity with AWGN $C(\beta = 2,\alpha,\mu)$. This integral was calculated in [13], and is given by eq. (12), where $\MeijerG{m}{n}{p}{q}{a_1,\ldots,a_p}{b_1,\ldots,b_q}{z}$ is the Meijer-G function [14, Sec. 7.8], $\rho = \frac{\mathbb{E}\{h\} \Gamma(\mu)}{\Gamma(\mu + \frac{2}{\alpha})}$, $\Phi(a,b) \equiv \frac{b}{a}, \frac{b+1}{a},...,\frac{b+a-1}{a}$, where $b$ is a positive integer and $a$ is an arbitrary real value, then $k$ is a positive integer that makes $\frac{\alpha k}{2}$a positive integer for some value of $\alpha$. The bounds on the ergodic capacity of the UWA is given in eq. (13).  

\begin{figure*}[!t]
\normalsize
\setcounter{mytempeqncnt}{\value{equation}}
\setcounter{equation}{10}
\begin{align}
\label{10}
C(2,\alpha,\mu) &= \int_{h = 0}^{\infty} \log\left(1+\frac{P h^{2}}{\sigma_{N}^{2}}\right) \times \frac{\alpha \mu^{\mu} h^{\alpha \mu-1}}{\tilde{h}^{\alpha \mu}\Gamma(\mu)} \mbox{exp}\left(-\mu \frac{h^{\alpha}}{\tilde{h}^{\alpha}} \right) dh \\
&= \frac{1}{2\log(2) \sqrt{k} \rho^{\frac{\alpha \mu}{2}} \Gamma(\mu) (2 \pi)^{\frac{k+\alpha k-3}{2}} } \times \MeijerG{k(\alpha+1)}{\frac{\alpha k}{2}}{\alpha k}{k(\alpha+1)}{\Phi(\frac{\alpha k}{2}, \frac{-\alpha \mu}{2}), \Phi(\frac{\alpha k}{2}, 1-\frac{\alpha \mu}{2})}{\Phi(k, 0), \Phi(\frac{\alpha k}{2}, \frac{-\alpha \mu}{2}), \Phi(\frac{\alpha k}{2}, \frac{-\alpha \mu}{2})}{\frac{\rho^{-\frac{\alpha k}{2}}}{k^{k}}}
\end{align}
\setcounter{equation}{\value{mytempeqncnt}+2}
\hrulefill
\vspace*{4pt}
\end{figure*}

\begin{figure*}[!t]
\normalsize
\setcounter{mytempeqncnt}{\value{equation}}
\setcounter{equation}{12}
\begin{align}
\label{10}
C(2,\alpha,\mu) \leq C(\beta,\alpha,\mu) \leq C(2,\alpha,\mu) + \frac{1}{2} \log \left(\frac{\beta^{2} \pi e^{1-\frac{2}{\beta}} \Gamma(\frac{3}{\beta})}{2(\Gamma(\frac{1}{\beta}))^{3}} \right).
\end{align}
\setcounter{equation}{\value{mytempeqncnt}+2}
\hrulefill
\vspace*{4pt}
\end{figure*}

\section{Underwater Secrecy Rates}

For a point-to-point UWA channel with the existence of one eavesdropper, the secrecy capacity $C_{s}$ in the conventional AWGN case can be given by
\begin{equation}
C_{s, AWGN} = \left\{\frac{1}{2} \log(1+SNR_{SD})-\frac{1}{2} \log(1+SNR_{SE})\right\}^{+},
\end{equation}
where $SNR_{SD}$ and $SNR_{SE}$ are the SNRs for the source-destination and source-eavesdropper links respectively, and ${x}^{+} = \max(0,x)$. The condition for the existence of a secrecy rate in the AWGN channel is $SNR_{SD} > SNR_{SE}$, i.e. the legitimate channel average SNR must be greater than that of the eavesdropper channel. For the UWA AWGGN, the secrecy capacity is given by eq. (14), where $\beta_{SD}$ and $\beta_{SE}$ are the shaping parameters of the noise perceived at the destination and eavesdropper receivers. In this case, the condition on the existence of a secrecy rate is given in eq. (15). The condition in eq. (15) suggests that a secrecy rate exists in the UWA channel even if $SNR_{SD} < SNR_{SE}$. In practice, this corresponds to a scenario where both legitimate transceivers exists at a shallow water level, while the eavesdropper is at a deeper level, i.e. legitimate and eavesdropper channels have different shaping parameters.  

\begin{figure*}[!t]
\normalsize
\setcounter{mytempeqncnt}{\value{equation}}
\setcounter{equation}{13}
\begin{equation}
 C_{s, AWGGN} = \left\{\frac{1}{2} \log(1+SNR_{SD}) + f(\beta_{SD})-\frac{1}{2} \log(1+SNR_{SE})-f(\beta_{SE})\right\}^{+}.
\end{equation}
\begin{equation}
\frac{\beta_{SD}^{2} e^{1-\frac{1}{\beta_{SD}}} \Gamma(\frac{3}{\beta_{SD}})}{\left(\Gamma(\frac{1}{\beta_{SD}})\right)^{3}} \times (1+SNR_{SD}) > \frac{\beta_{SE}^{2} e^{1-\frac{1}{\beta_{SE}}} \Gamma(\frac{3}{\beta_{SE}})}{\left(\Gamma(\frac{1}{\beta_{SE}})\right)^{3}} \times (1+SNR_{SE})
\end{equation}
\setcounter{equation}{\value{mytempeqncnt}+2}
\hrulefill
\vspace*{4pt}
\end{figure*}

\section{Numerical results}

Figure 3 depicts the upper bound capacity in eq. (10) versus SNR for different values of $\beta$, we note that at lower values of SNR the $C_{AWGN}$ term will be very small while the f($\beta$) term will become dominant and the difference appears here between different curves, but at larger values of SNR the $C_{AWGN}$ term will be very large and dominate leading to less difference between curves.

Figure 4 depicts the upper bound capacity for the case of $\alpha$-$\mu$ fading in eq. (13) versus SNR for different values of $\alpha$, we note that for any SNR and for a given value of $\beta$ the upper bound capacity increases with $\alpha$.While for fixed $\alpha$ and $\mu$ the only difference that changing the value of $\beta$ makes is that it adds a fixed value of g($\beta$) to the curve.

Figure 5 depicts the Secrecy rate versus $SNR_{SD}$ for two cases. The first case is when $\beta_{SD}=\beta_{SE}=2$, which is the normal AWGN channel case, and $SNR_{SE}=-5$, it is shown that secrecy rate is positive only when $SNR_{SD}>-5$, which is predicted according to eq. (16). The second case is when $\beta_{SD}=1.5$, $\beta_{SE}=0.8$ and $SNR_{SE}=-5$, it is found that secrecy rate exists even at $SNR_{SD}<-5$, this is because $\beta_{SE}$ and $SNR_{SE}$ have different values now so eq. (15) can be satisfied even if $SNR_{SD}<SNR_{SE}$.

\begin{figure}[!t]
\centering
\includegraphics[width=3.25in]{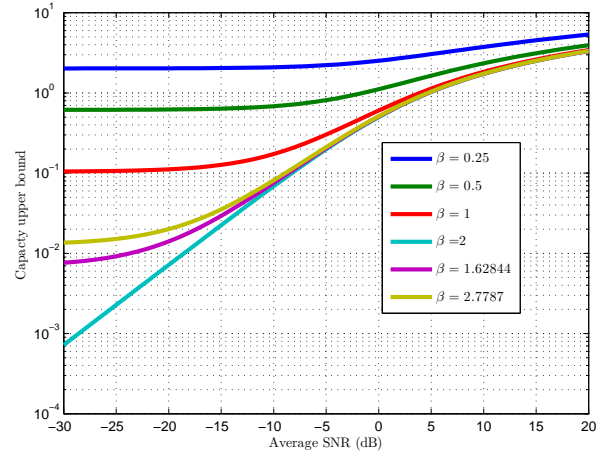}
\caption{The Capacity upper bound of AWGGN channel for different values of $\beta$.}
\label{fig_sim}
\end{figure}

\begin{figure}[!t]
\centering
\includegraphics[width=3.25in]{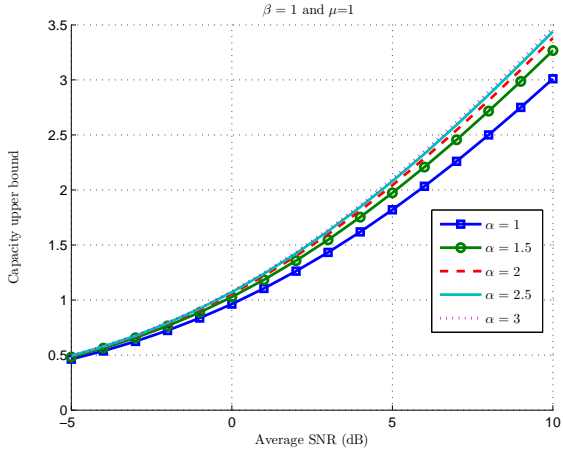}
\caption{The Capacity upper bound of AWGGN channel with $\alpha$-$\mu$ fading for different values of $\alpha$ and $\beta$=1.}
\label{fig_sim}
\end{figure}

\begin{figure}[!t]
\centering
\includegraphics[width=3.25in]{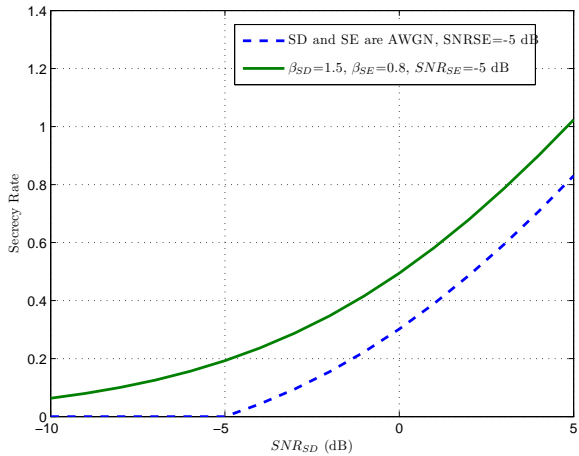}
\caption{The Secrecy rate for two different cases of S-D and S-E channels.}
\label{fig_sim}
\end{figure}

\section{Conclusions}
Based on previous works that proved that underwater acoustic channels are characterized by Generalized Gaussian noise and $\alpha-\mu$ fading, we derive in this paper an upper bound for capacity of underwater acoustic channel with AWGGN and $\alpha-\mu$ fading. We found that a function of the shaping parameter $\beta$ represents the increase in $C_{AWGGN}$ with respect to $C_{AWGN}$. This function was found to have much larger values when $\beta$ has values lower than 1.We studied the secrecy rate of a system where the Source-Destination and the Source-eavesdropper channels are characterized by AWGGN, we derived the condition on having secrecy rate$>0$ it was found that the condition depends on $\beta_{SD} and beta_{SE}$ beside $SNR_{SD} and SNR_{SE}$, so we may have $SNR_{SD}<SNR_{SE}$ and still have secrecy rate.


\begin{thebibliography}{1}
\bibitem{1}
D.~Pompili, I.~.F.~Akyildiz, and T.~Melodia, ``Challenges for efficient communication in underwater acoustic sensor networks," \emph{ACM SIGBED Review}, vol. 1, no. 2, pp. 3-8, July 2004.
\bibitem{31}
W.-Y.~Shin, D.~E.~Lucani, M.~Médard, M.~Stojanovic, V.~Tarokh,``On the effects of frequency scaling over capacity scaling in underwater networks—Part II: Dense network model," \emph{Wireless personal communications}, Vol. 71, pp. 1701-1719, 2013.
\bibitem{32}
D.~E.~Lucani, M.~Médard, M.~Stojanovic,``Capacity Scaling Laws for Underwater Networks," \emph{Internet Mathematics,} pp. 241-264, 2013.
\bibitem{33}
C.R. Berger, S. Zhou, J.C. Preisig, and P. Willett,``Sparse Channel Estimation for Multicarrier Underwater Acoustic Communication: From Subspace Methods to Compressed Sensing," \emph{IEEE Transactions on Signal Processing}, Vol. 58, pp. 1708 - 1721, March 2010.
\bibitem{3}
S.~Banerjee and M.~Agrawa,``Underwater acoustic noise with generalized Gaussian statistics: Effects on error performance," \emph{2013 MTS/IEEE, Bergen, Norway,} pp. 1-8, june 2007.
\bibitem{5}
H.~Soury, F.~Yilmaz, and M.~S.~Alouini,``Average Bit Error Probability of Binary Coherent Signaling over Generalized Fading Channels Subject to Additive Generalized Gaussian Noise," \emph{IEEE Communications Letters},vol. 16, no. 6, pp. 785-788, April 2012.
\bibitem{6}
D. E. Lucani, M. Médard, and M. Stojanovic, ``Capacity scaling laws for underwater networks," \emph{IEEE 42nd Asilomar Conference on Signals, Systems and Computers, Pacific Grove, CA, US} pp. 2125 - 2129, Oct. 2008.
\bibitem{9}
L. Matos, G. Ferreira, J. Panaro, L. Barreira, and E. Mainetti``Time and Frequency Fading Statistics for Underwater Acoustic Signals in Shallow Water," \emph{NATO Oceanic Engineering}.
\bibitem{11}
S. Nadarajah,`` A generalized normal distribution,"\emph{Journal of Applied Statistics,} vol. 32, no. 7, pp. 685 - 694, April. 2011.
\bibitem{13}
T. M. Cover and J. A. Thomas, ``Elements of information theory," John Wiley and Sons, 2012.
\bibitem{15}
S. N. Diggavi, and T. M. Cover, ``The worst additive noise under a covariance constraint," \emph{ IEEE Transactions on Information Theory,} pp.3072 - 3081, Nov. 2001.
\bibitem{16}
I. Shomorony, and S. Avestimehr, ``Worst-Case Additive Noise in Wireless Networks," \emph{ IEEE Transactions on Information Theory,} pp.3833 - 3847, June 2013.
\bibitem{23}
D. B. da Costa and M. D. Yacoub,``Average channel capacity for generalized fading scenarios," \emph{ IEEE Communications Letters,} vol. 11, no. 12, pp. 949 - 951, Dec. 2007.
\bibitem{20}
A. Jeffrey and D. Zwillinger, ``Table of Integrals, Series, and Products" \emph{Academic Press}, 2000.
\end{thebibliography}
\end{document}